\documentclass[reprint,aip,apl
]{revtex4-1}
\usepackage{docs}%
\usepackage{bm}%
\usepackage{epstopdf}
\usepackage{graphicx}
\usepackage{upgreek}
\expandafter\ifx\csname package@font\endcsname\relax\else
 \expandafter\expandafter
 \expandafter\usepackage
 \expandafter\expandafter
 \expandafter{\csname package@font\endcsname}%
\fi
\begin{document}

\title{Intrinsic Ultrathin Topological Insulators Grown via MBE Characterized by in-situ Angle Resolved Photoemission Spectroscopy}%

\author{J. J. Lee}%
\affiliation{Department of Applied Physics, Stanford University, Stanford, California 94305, USA}
\affiliation{Stanford Institute for Materials and Energy Sciences, SLAC National Accelerator Laboratory, 2575 Sand Hill Road, Menlo Park, California 94025, USA}
\affiliation{Geballe Laboratory for Advanced Materials, Department of Applied Physics, Stanford University, Stanford, California 94305, USA}
\author{F. T. Schmitt}
\author{R. G. Moore}
\affiliation{Stanford Institute for Materials and Energy Sciences, SLAC National Accelerator Laboratory, 2575 Sand Hill Road, Menlo Park, California 94025, USA}
\affiliation{Geballe Laboratory for Advanced Materials, Department of Applied Physics, Stanford University, Stanford, California 94305, USA}
\author{I. M. Vishik}
\author{Y. Ma}
\affiliation{Department of Applied Physics, Stanford University, Stanford, California 94305, USA}
\affiliation{Stanford Institute for Materials and Energy Sciences, SLAC National Accelerator Laboratory, 2575 Sand Hill Road, Menlo Park, California 94025, USA}
\affiliation{Geballe Laboratory for Advanced Materials, Department of Applied Physics, Stanford University, Stanford, California 94305, USA}
\author{Z. X. Shen}
\affiliation{Stanford Institute for Materials and Energy Sciences, SLAC National Accelerator Laboratory, 2575 Sand Hill Road, Menlo Park, California 94025, USA}
\affiliation{Geballe Laboratory for Advanced Materials, Department of Applied Physics, Stanford University, Stanford, California 94305, USA}

\date{\today}
\begin{abstract}
We demonstrate the capability of growing high quality ultrathin (QL $\lesssim$ 10) films of the topological insulators Bi$_{2}$Se$_{3}$ and Bi$_{2}$Te$_{3}$ using molecular beam epitaxy. Unlike previous growth techniques, which often pin the Fermi energy in the conduction band for ultrathin samples, our samples remain intrinsic bulk insulators. We characterize these films using in-situ angle resolved photoemission spectroscopy (ARPES), which is a direct probe of bandstructure, and ex-situ atomic force microscopy. We find that the conduction band lies above the Fermi energy, indicating bulk insulating behavior with only the surface states crossing E$_{F}$. We conclude that thermal cracking of Te and Se in our growth leads to higher quality thin films, paving the way for future improvements in growth of topological insulators.
\end{abstract}

\maketitle

Topological insulators are a recently discovered class of materials exhibiting unique surface states which are strongly protected against perturbations in either the material bulk or on the surface \cite{ RevModPhys.82.3045, RevModPhys.83.1057, PhysRevLett.98.106803, ISI:000266544800016, Chen10072009, ISI:000266544800024, ISI:000285501100012, PhysRevLett.103.146401}. These states are characterized by a linear electronic dispersion where the electron's spin is locked to its momentum direction. The surface states expand across the bulk band gap and intersect to form a Dirac cone centered around $\Gamma$. Ideally, the Fermi level will cross the surface states inside the bulk band gap, making the material a surface metal. Experimentally, however, such materials often have the Fermi energy pinned into the bulk conduction band or have a conduction surface state coexist with the topological surface state\cite{Chen10072009,ISI:000266544800016, ISI:000288224800026}. For the parent compounds Bi$_{2}$Se$_{3}$ and Bi$_{2}$Te$_{3}$, such pinning of the Fermi energy has been attributed to bulk crystal defects\cite{PhysRevB.79.195208}. Thus, an important experimental challenge has been to reduce the contribution of bulk carriers such that the surface states dominate transport properties. \\

Because Bi$_{2}$Se$_{3}$ and Bi$_{2}$Te$_{3}$ possess a layered rhombohedral structure, a large effort has been focused on thin film growth techniques such as molecular beam epitaxy to reduce crystal defects\cite{zhang:053114, ADMA:ADMA201100678, Krumrain2011115, Bansal2011224}. Most thin film growth previously reported have made use of thermal effusion cells. However, when selenium or tellurium are evaporated in vacuum, they form molecular complexes of variable atomic number, most commonly Te$_{2}$/Se$_{2}$ \cite{ADMA:ADMA201000368, Krumrain2011115,Bansal2011224}. This often requires a larger Se$_{2}$/Te$_{2}$ flux be added to compensate, often in a ratio greater than 10 Te$_{2}$/Se$_{2}$ to 1 Bi\cite{Bansal2011224, Krumrain2011115}.  In addition, such growth methods still tend to pin the Fermi energy to some level above the conduction band, thus reducing the contribution of the topological surface state to transport\cite{ADMA:ADMA201000368, ISI:000281540200024}. Traditionally, dopants have been added to the crystal growth to tune the Fermi energy of the system into the bulk gap\cite{ISI:000299921000034, PhysRevB.84.165311, ISI:000299159900006, ISI:000269314000032}.  This letter reports on recent advances in growth of intrinsic topological insulator parent compounds Bi$_{2}$Se$_{3}$ and Bi$_{2}$Te$_{3}$ using a selenium and tellurium thermal cracker effusion cell, which to the authors' knowledge, has not been explicitly reported in previous works. Combined with high energy resolution, in-situ angle-resolved photoemission spectroscopy, we demonstrate that films grown this way have their Fermi energy located within the bulk gap.\\

\begin{figure}[b]
\includegraphics[scale = .4]{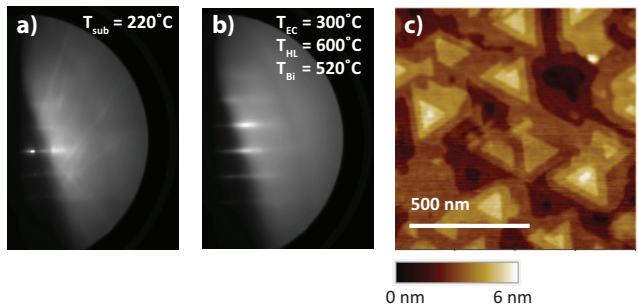}
\caption{Growth of Bi$_{2}$Te$_{3}$. Figure a) shows the reflection high-energy electron diffraction spots of the sapphire substrate before growth. Fig b) shows the RHEED pattern 250 seconds after shutters are opened. The formation of streaks indicates epitaxial growth. Fig c) shows the ex-situ AFM of the sample. The topmost 2 layers have island growth but below that on sees nearly full coverage. }
\end{figure}

Figure 1 shows the growth conditions of Bi$_{2}$Te$_{3}$. The substrate used was a sapphire-0001 substrate cleaned for 1.5 hours at 650$^{\circ}$, followed by 30 minutes at 900$^{\circ}$ C.  The effusion cell used during growth contains a hot lip with a cracking insert which helps break Te$_{2}$/Se$_{2}$ molecular flux into atomic Te/Se.  We grow with flux ratios that are much more stoichiometric than previous groups (2 Te/Se to 1 Bi). We find that growth rate is controlled via the bismuth flux, as reported in ref \cite{Krumrain2011115,ADMA:ADMA201000368}. However, exceedingly large Te fluxes produced lower quality growth, manifested by broader and weaker surface dispersion spectra. Figure 1(a) shows reflection high-energy electron diffraction pattern of the bare sapphire substrate. Figure 1(b) shows the RHEED pattern after several minutes of growth. The streak intensities show clear oscillations (not shown), indicative of epitaxial growth, with an oscillation period of approximately 50 seconds. Bi$_{2}$Te$_{3}$ films grown for this study ranged from 5 to 10 QL, with similar electronic structure. Figure 1(c) shows the AFM image of a $\mathord{\sim}$10 QL sample taken ex-situ. Clear trianglular islands can be resolved in the topmost two layers, while the lower layers have nearly full coverage.\\

\begin{figure}[t]
\includegraphics[scale = .5]{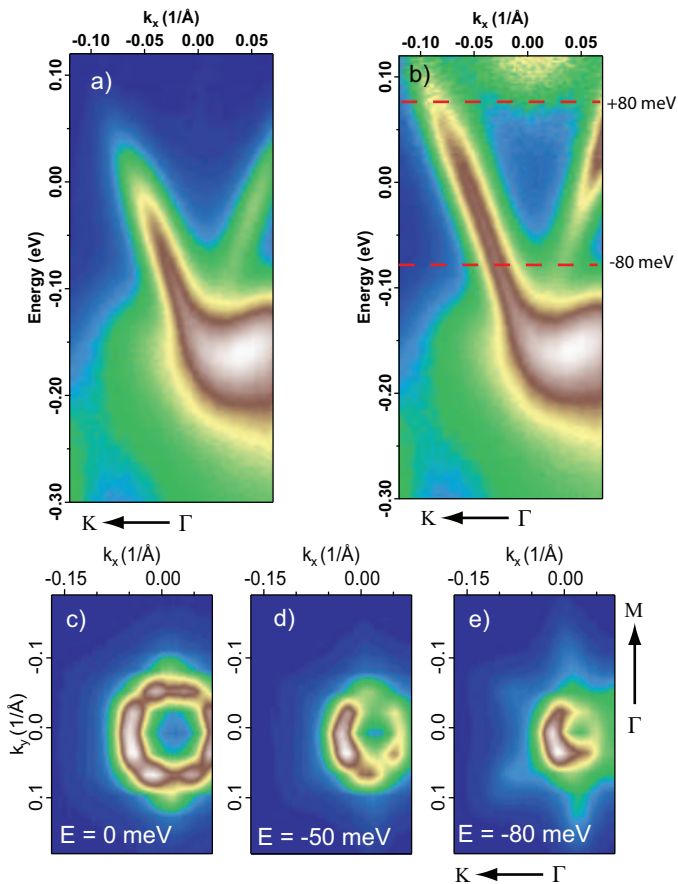}
\caption{ARPES spectra of $\mathord{\sim}$10 QL layers bismuth telluride. Fig a) is a dispersion along $\Gamma$. The Dirac cone can be clearly resolved connecting with the  bulk valence band. Fig b) is the same dispersion with the Fermi-Dirac function factored out, where we can see the thermally populated conduction band. The dotted red lines indicate the location of the valence band maximum and conduction band minimum. Figures c) - e) show constant energy mappings, with energies stated relative to the Fermi energy. At 80 meV below the Fermi energy, we see the appearance of the valence band.}
\end{figure}

Electronic properties of such thin films are often studied ex-situ through transport or quantum oscillations\cite{PhysRevB.83.165440}. However, to determine the Fermi level and obtain bandstructure directly, we use angle resolved photoemission spectroscopy (ARPES) \cite{RevModPhys.75.473}. Due to ARPES's extreme surface sensitivity, such measurements are ideally performed in situ. Previous in-situ ARPES measurements on ultrathin films, whose thicknesses are less than 20-30 quintuple layers, have revealed a bulk conduction band or quantum well states\cite{ADMA:ADMA201000368, ISI:000281540200024, plucinski:222503, Chang.Spin.BiSeonSapphire}. These quantum well states are due to confinement along the \textit{c}-axis in films of thicknesses around 5 QL. In our setup, a Scienta R2002 electron analyzer was used with a 7 eV laser light source. Fermi energies were determined using a gold spectra, with $<$5 meV resolution. Figure 2(a) shows the electron dispersion along a high symmetry axis from $\Gamma$ to K, taken 4 hours after growth, at room temperature. The Dirac cone can be clearly distinguished, intersecting with the valence band. Noticeably there is no signature of the conduction band. To determine the location of the conduction band, we use thermal population of states above the Fermi energy. Figure 2(b) plots the spectra of 2(a) after factoring out the Fermi-Dirac function at room temperature. We can thus resolve the bulk conduction band to be approximately 80 meV above the Fermi energy. Furthermore we can determine the location of the valence band through mapping the entire 2D bandstructure and examining the appearance of the valence band away from the gamma point. Figures 2(c)-(e) are such mappings at different binding energies. 2(c), at 0 meV, shows the Fermi surface, which reveals no trace of the bulk conduction band state often seen in other materials grown by other methods. In 2(d), which is 50 meV below the Fermi energy, one can see the formation of the Dirac cone. Finally, in figure 2(e), at 80 meV below the fermi energy, one sees the appearance of the hexagonal bulk valence band state\cite{Chen10072009}. Our data thus gives a bulk gap of approximately 160 meV, from valence band maximum to conduction band minimum. Such a gap is consistent with previous experimental results at other photon energies\cite{Chen10072009,ADMA:ADMA201000368}. We therefore conclude our Fermi energy lies exactly midgap. \\

\begin{figure}[t]
\includegraphics[scale = .5]{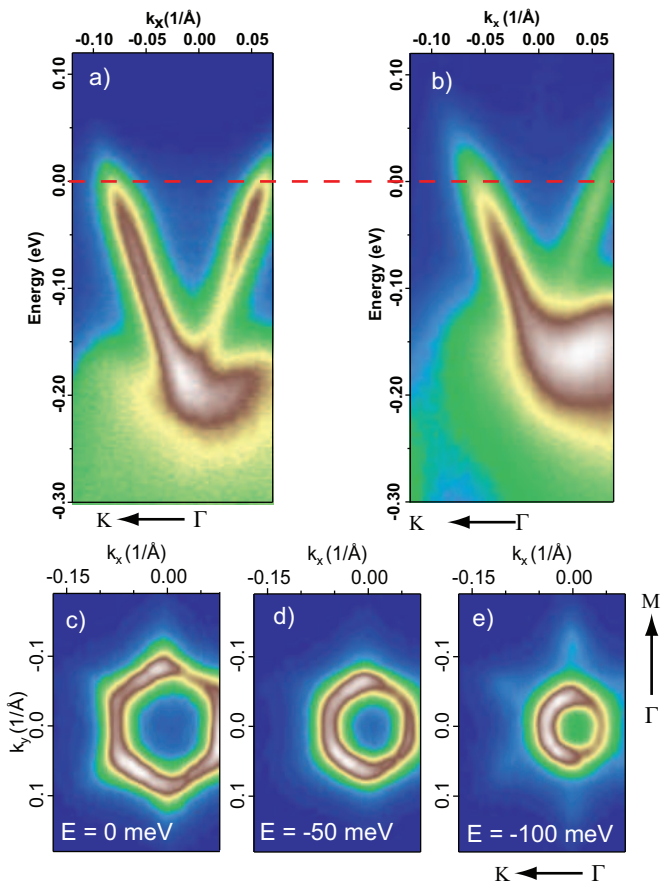}
\caption{ARPES spectra of bismuth telluride at 150K and after 24 hour in UHV. Figure a) shows a dispersion through $\Gamma$. Figure b) shows the same dispersion for the earlier high temperature data. The red dotted line is given to show the band shift. We see the dirac point having shifted by 20 meV lower after aging. Figures c) - e) show constant energy mappings at different binding energies below the Fermi surface. The band shifting is confirmed through the increase Fermi surface area in figure c). We see the emergence of the valence band at 100 meV below the Fermi energy.}
\end{figure}

It has been previously reported that Bi$_{2}$Te$_{3}$ exhibits a downward band-bending as one decreases temperature\cite{ADMA:ADMA201000368,1367-2630-13-1-013008}. To measure the temperature dependence we map the bandstructure at 150K. Data were taken approximately 24 hours after growth. Figure 3(a) plots the dispersion from $\Gamma$ to K. 3(b) is a reproduction of 2(b) for comparison. Figures 3(c)-(e) show constant energy mappings. The fermi surface has attained a noticeable hexagonal shape, consistent with previous low temperature measurements \cite{Chen06032012, Chen10072009}. Scanning through constant-energy maps with increasing binding energy, we see the appearance of the valence band at approximately 100 meV below the surface, shown in figure 3(e). This indicates a band shift of approximately 20 meV from 300K to 150K. Due to measuring at lower temperature, we cannot use thermal population to observe the conduction band. The momentum distribution curve linewidths do not broaden, indicating little to no aging, which has otherwise been previously reported in such bulk systems \cite{Chen06032012,PhysRevLett.103.146401}. Overall, the effects of aging in vacuum and temperature variation do not affect the Fermi level to the point where our film is degenerately doped. In addition, we find that variations in substrate temperatures on the order of 40 degrees do not send the Fermi level outside the gap.\\

\begin{figure}[t]
\includegraphics[scale = .5]{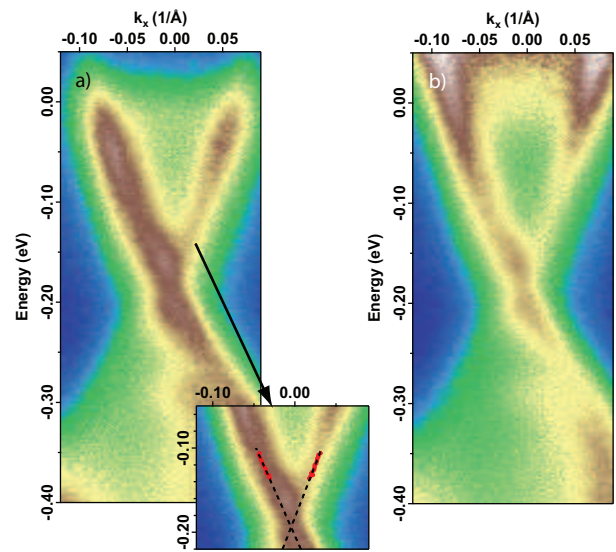}
\caption{ARPES spectra of bismuth selenide. Figure a) shows the dispersion along $\Gamma$. The inset shows the fitting used to extract the dirac point, which is located approximately 190 meV below the Fermi surface. Figure b) shows the dispersion with the Fermi-Dirac function factored out. The increase in surface state intensity is indicative of scattering towards the conduction band.}
\end{figure}

In addition to growing ultrathin films of Bi$_{2}$Te$_{3}$, we also grow Bi$_{2}$Se$_{3}$ using a thermal cracker cell. Following similar procedures and flux ratios, and using the same substrate, we grow films with thicknesses ranging from 6-8 QL. ARPES spectra of an 8 QL film are plotted in figure 4. Figure 4(a) shows the dispersion along the $\Gamma$-K axis at 250K. To determine the Dirac point, we fit the momentum distribution peaks of the surface states to a linear dispersion near the dirac point (inset of figure 4(a)). The intersection of the linear dispersions determines the Dirac point. We find the Dirac point lies approximately 190 meV below the Fermi energy. Previous work on bulk samples has shown the conduction band to lie 200-220 meV above the Dirac point \cite{ISI:000266544800016, ISI:000288224800026, 1367-2630-13-1-013008, PhysRevB.81.041405}, again above our Fermi energy.  Figure 4(b) shows the spectrum after factoring out the Fermi-Dirac function. Using this normalized spectrum we find at energies 10-20 meV above E$_{F}$ an increase in linewidth of the surface states bands. This is indicative of surface state to bulk state scattering \cite{1367-2630-13-1-013008}, and hence gives a signature of the bulk band bottom. Hence surface conduction from non-topological surface states is dramatically reduced. Given a temperature low enough to avoid thermal population, one would expect bulk conduction to be small. Other films grown in this study exhibited similar electronic structure.\\

In summary, we have demonstrated the ability to grow high quality ultra-thin films of the topological insulators Bi$_{2}$Se$_{3}$ and Bi$_{2}$Te$_{3}$ using a thermal cracker effusion cell. With in-situ high resolution ARPES, we characterized the quality of the films immediately after growth by mapping the band structure.  For our undoped, as grown films of Bi$_{2}$Te$_{3}$, the Fermi energy was always located well-within the bulk gap. Lowering of temperature causes a rigid shift of the bands towards n-type by approximately 20 meV, but leaves the overall band structure, doping and TI properties intact. For Bi$_{2}$Se$_{3}$, growth of films show the conduction band to be 10-20 meV above E$_{F}$. The use of thermal cracker cells to grow our materials should pave the way for better materials growth of topological insulators, and opens up a wide range of possibilities not only scientifically but also from an applied perspective, since ultrathin films of sufficient quality and stability for surface-dominated transport can be fabricated for further study.\\

This work is supported by DOE office of Basic Energy Science, Division of Materials Science.

\bibliographystyle{apsrev4-1}
\bibliography{ref_mbeTI}
\end{document}